\documentclass[preprint,showpacs,preprintnumbers,amsmath,amssymb]{revtex4}

\usepackage{graphicx}% Include figure files

\begin{document}

\title{Modelling chemical reactions using semiconductor quantum dots}
\author{ A. Yu. Smirnov$^{1,2,3}$, S. Savel'ev$^{1,4}$,  L. G. Mourokh$^{1,3,5,6}$, and Franco Nori$^{1,7}$}

\affiliation{ $^1$ Frontier Research System, The Institute of Physical and
Chemical Research (RIKEN), Wako-shi, Saitama, 351-0198, Japan \\
$^2$ CREST, Japan Science and Technology Agency, Kawaguchi,
Saitama, 332-0012, Japan \\
$^3$ Quantum
Cat Analytics, 1751 67 St. E11, Brooklyn, New York 11204, USA \\
$^4$ Department of Physics, Loughborough University, Loughborough
LE11 3TU, UK \\
$^5$ Department of Physics, Queens College, The City University of
New York, Flushing, New York 11367, USA \\
$^6$ Department of Engineering Science and Physics, College of Staten Island, The City University of
New York, Staten Island, New York 10314, USA \\
$^7$ Center for Theoretical Physics, Physics Department, The University of Michigan, Ann Arbor, MI 48109-1040, USA}

\date{\today}

\begin{abstract}
{We propose using semiconductor quantum dots for a simulation of chemical reactions as electrons are redistributed among such
artificial atoms. We show that it is possible to achieve various reaction regimes and obtain different reaction products by
varying the speed of voltage changes applied to the gates forming quantum dots. Considering the simplest possible reaction,
$H_2+H\rightarrow H+H_2$, we show how the necessary initial state can be obtained and what voltage pulses should be applied to
achieve a desirable final product. Our calculations have been performed using the Pechukas gas approach, which can be extended for
more complicated reactions.}
\end{abstract}

\pacs{73.21.La, 85.35.Be}

\maketitle

Detailed simulations of chemical and biological processes can provide crucial insight on these and help determining optimal
experimental regimes and conditions. However, the high-accuracy modelling, at the quantum level, of even the simplest chemical
reactions represents a significant challenge because it encompasses changes that involve the motion of electrons in the forming
and breaking of chemical bonds. On classical computers, the resource requirements for the complete simulation of the
time-dependent Schr\"odinger equation scale exponentially with the number of atoms in a molecule, imposing very severe limitations
in the systems that can be modelled. However, recent developments of novel quantum computation schemes allow a polynomial scale of
required resources. Via these approaches, a quantum system can simulate the behavior of another quantum system of interest (see,
e.g. \cite{Lloyd1,Manousakis1,AspuruGuzik1}).

Semiconductor quantum dots can be described as artificial atoms (see, e.g. \cite{1DExp}). These have discrete electron spectra
revealing a shell structure and exchange corrections to the electron energies according to Hund's rules. In this sense, {\it
coupled quantum dots} can be regarded as {\it artificial molecules} \cite{Kouwenhoven2}. Depending on the tunnel coupling
strengths, electron distribution, and shell structure, the dots can form both ionic- and covalent-like bonds. Manifestations of
these molecular states in double-dot structures were observed by numerous groups \cite{DDExp}. The idea of using the charge
degrees of freedom in double-dot systems as a qubit has been proposed theoretically \cite{Barenco1} and implemented experimentally
\cite{Fujisawa1}.

Recent achievements in nanotechnology facilitate the precise control of the number of electrons in quantum dots and the tunnel
energy splittings, by tuning the voltages applied to the gates \cite{SEExp}. Measuring the current through a quantum point contact
in the vicinity of the structure allows the determination of the exact charge locations \cite{Field1}. Moreover, structures with
{\it three} coupled quantum dots have been recently fabricated and characterized \cite{TDExp,Louis06} with the potential to easily
increase the number of dots, as needed.

Based on these developments, we propose to employ the electron redistribution in coupled quantum dot systems for chemical reaction
modelling. The number of electrons in the first and second quantum dot shells are 2 and 4, respectively. Accordingly, a quantum
dot with one electron can be considered as an {\it artificial hydrogen atom} (one electron vacancy in the outer shell) and a
quantum dot containing four electrons can be viewed as an {\it artificial oxygen atom} (two electron vacancies in the outer
shell). Consequently, the coupling of these three dots (which can be easily controlled by changing the gates' potentials) can
model the covalent molecular bond formation between the four-electron dot and each of the one-electron dots. This would represent
the hydrogen oxidation reaction, with the formation of an {\it artificial water molecule}. Increasing the number of dots would
allow the modelling of more complicated reactions. Moreover, such {\it artificial chemical reactions} can be done under conditions
(such as the presence of an external magnetic field) not readily accessible in all real molecules. Furthermore, the speed of the
reactions could be easily varied in a very wide range. In quantum chemistry, calculations of chemical reactions usually employ the
molecular Hamiltonian written in the second-quantized form \cite{AspuruGuzik1},
\begin{equation}
H = \sum_{pq}\langle p|H_0|q\rangle ~ a_p^+a_q - \frac{1}{2}\sum_{pqrs}\langle pq|V_e|rs\rangle ~a_p^+a_q^+a_ra_s. \label{Ham1}
\end{equation}
Here $a_p^+~(a_p)$ are Fermi operators responsible for a creation (annihilation) of an electron in a single-particle orbital
$|p\rangle$, $|pq\rangle = |p\rangle \otimes |q\rangle $ is a two-electron state, $H_0$ is a single-particle Hamiltonian
consisting of kinetic and nuclear attraction operators, and $V_e$ is a term related to the electron-electron repulsion. The system
of coupled quantum dots is characterized by a similar Hamiltonian. In the first step of the modelling, each single-particle atomic
orbital of the molecules should be mapped into a single-particle orbital of the quantum dot system. Only the active states
participating in the reaction must be selected. Afterwards we have to carefully choose control parameters for the dots (gate
voltages, barriers heights, distances between dots, magnetic fields, etc.) with the aim of mapping the energy spectrum of the real
molecules to the spectrum of the dots (with a fixed scale coefficient). The possibility to do this efficiently is supported by
recent experiments \cite{Louis06} where four quantum states of three dissimilar semiconductor quantum dots were tuned in resonance
to form multiple quadruple points on the stability diagram, thus demonstrating the fine tunability that quantum dot structures can
achieve. Therefore, even though, contrary to three-dimensional atoms, the dots are quasi-two-dimensional objects, we believe that
the proposed approach can qualitatively and sometimes quantitatively describe the outcome of real chemical reactions.

The Hamiltonian of the system under study can be separated into two parts
\begin{equation}
H(t) = H_0 + \lambda(t)V\label{eq_H0}
\end{equation}
where $H_0$ describes the invariant part and $\lambda(t)V$ is responsible for the reaction. The time dependence of the parameter
$\lambda(t)$ can be chosen specifically for various reaction regimes. For small $\dot{\lambda}\equiv d\lambda /dt$ we have an
adiabatic evolution, with the system following its ground state during the reaction. For extremely large $\dot{\lambda}$, the
system's state remains unchanged; and for the intermediate case, several Landau-Zener transitions may occur at the avoided
crossing points, with various states being populated after the reaction. It should be noted that even the "slow" evolution
mentioned above must be faster than the decoherence time for the coupled-quantum-dots system, which is about 1 ns according to
Ref. \cite{Fujisawa1}. In contrast to adiabatic quantum computing \cite{aqc,Zagoskin1}, where the aim is to keep the system either
at or near its ground state, here we focus on a completely different issue: {\it how to control the population of desirable (not
necessarily ground) states by changing the speed of evolution and the shape of $\lambda(t)$}. Harnessing the constructive features
of the Landau-Zener effect allows us to travel more effectively in the whole Hilbert space, not only near its bottom part. From
this point of view, the proposed approach can be considered as a step towards a better control of the quantum-mechanical state of
the system, which is one of the broad objectives of the quantum information processing.

The evolution of the instantaneous energy levels $E_n(\lambda)$ and eigenfunctions $|n\rangle$ of Hamiltonian (\ref{eq_H0}) can be
exactly mapped \cite{Zagoskin1} on the classical Hamiltonian dynamics of a 1D gas of fictitious particles (Pechukas gas
\cite{Pechukas1983}), with positions $x_n(\lambda) = E_n(\lambda)$ and momenta $v_n(\lambda) = V_{nn}(\lambda)$. The ``particle
repulsion'' is determined by the additional set of variables, the ``angular moments'' $l_{mn}(\lambda) =
[E_m(\lambda)-E_n(\lambda)]V_{mn}(\lambda)$:
  \begin{eqnarray}\label{eq_Pechukas1}
\frac{d}{d\lambda}{x}_m&=&v_m   \nonumber \\
\frac{d}{d\lambda}v_m&=& 2\sum_{m\ne
n}\frac{|l_{mn}|^2}{(x_m-x_n)^3} \\
\frac{d}{d\lambda}{l}_{mn}&=& \sum_{k\neq m,n}   l_{mk}l_{kn} \left(\frac{1}{(x_m-x_k)^2}-\frac{1}{(x_k-x_n)^2}\right). \nonumber
\end{eqnarray}
Note that all the matrix elements in Eq.~(\ref{eq_Pechukas1}) are taken between the instantaneous eigenstates of the  Hamiltonian
(\ref{eq_H0}). The probabilities of the Landau-Zener transitions between states $m\leftrightarrow n$ is given by
\begin{equation}
p_{m,n}\;=\;\exp\left(-\frac{(\Delta_{\min}^{m,n})^3}{4\pi\hbar l_{mn}|\dot{\lambda}|}\right), \label{Pechukas2}
\end{equation}
where $\Delta_{min}^{m,n}$ is the minimal separation of levels at avoided crossing. We chose to use the Pechukas gas approach
because of its potential scalability for systems with a large number of elements \cite{Zagoskin1}.

Here, we examine the simplest possible chemical reaction: the scattering of a hydrogen atom from a hydrogen molecule ($H + H_2
\rightarrow H_2 + H$), see Fig.~1. Even though this reaction was performed and theoretically described for nearly a century, some
details (not understandable without accurate modelling at the quantum level) were observed recently \cite{HH2} (where the slightly
different reaction, $H + D_2 \rightarrow HD + D$, was studied). Although the reaction was dominated by a direct recoil mechanism
(when the incident hydrogen atom recoils along its original path after removing a deuterium atom to form a HD molecule), a second
slower reaction mechanism occurs with a time delay of 25 fs. One of the possible explanations of such time delay is the formation
of a metastable ``quasi-bound'' quantum state decaying into the reaction products. Such a system, with three nuclei and three
electrons, can be mapped onto the triple-quantum-dot system with the Hamiltonian
\begin{equation}
H= H_{3D} + H_C + H_{{\rm tun}} , \label{Htot}
\end{equation}
where
\begin{eqnarray}
H_{3D} &=& \sum_{S=1,2} \left( E_{AS}N_{AS} + E_{BS}N_{BS} + E_{CS}N_{C1S}\right)  , \nonumber \\
H_C &=& U_{A} N_{A1}N_{A2} + U_{B} N_{B1}N_{B2} + U_{C} N_{C1}N_{C2} + U_{AB} N_A N_B + U_{BC}N_B N_C + U_{AC}N_A N_C  ,
\nonumber \\
H_{{\rm tun}} &=& - \sum_{S=1,2} \left( \Delta_{AB}a_{AS}^+a_{BS} + \Delta_{BC}a_{BS}^+a_{CS} + \Delta_{AC}a_{AS}^+a_{CS} + h.c.
\right)  , \label{H3D}
\end{eqnarray}
$S=1(2)$ for spin-up(-down) electrons, $A,B,C$ are the dot indices and $N_{A,B,C}$ are the total populations of the corresponding
dots. The Hamiltonian, Eqs.~(\ref{Htot},\ref{H3D}), has 20 eigenfunctions and eigenvalues for the three-electron case, which can
be determined from the solution of the corresponding Schr\"odinger equation for specific values of the system parameters. It
should be noted that these parameters can be controlled by the gates' voltages applied to the triple-dot system. We will use the
following pulse sequence: in the first stage, the desired initial state is formed; in the second stage, an extremely fast
restoration pulse is used to return the gates into their initial conditions while simultaneously preserving the system state;
finally, in the third stage, the desired final state is obtained.

To better link to experiments, we choose the intradot Coulomb energies $U_A=U_B=U_C=2$ meV, the interdot Coulomb energies
$U_{AB}=U_{BC}=U_{AC}=0.2$ meV, and the tunnel matrix elements $\Delta_{AB}=\Delta_{BC}=\Delta_{AC}=0.05$ meV. We also introduce a
small Zeeman energy $E_1-E_2=0.003$ meV for all three dots, to lift the spin degeneracy. The dot energies before reaction
($\lambda=0$) are chosen as $E_B=E_C=E_A+0.5E_P$ with $E_P =2.2$ meV. In this case, the ground state for the Hamiltonian,
Eq.~(\ref{Htot}), is given by
\begin{equation}
\Psi_1 (\lambda=0) = \sqrt{{2\over 3}}\, a^+_{A1}a^+_{B2}a^+_{C2}
 |0 \rangle - \sqrt{{1\over 3}}\, {a^+_{B1}a^+_{C2} + a^+_{B2}a^+_{C1}\over\sqrt{2}}\times a^+_{A2}|0
 \rangle  , \label{Psi1}
\end{equation}
i.e., it is a superposition of the state $ a^+_{A1}a^+_{B2}a^+_{C2} |0 \rangle $, where a single electron is located in each dot
(no bonds, with a probability 2/3), and the state,a spin-triplet $ T_0$, formed in the dots B and C, plus one electron located on
the dot A (with a probability 1/3). It should be noted that the the real chemical bond is formed by the spin-singlet state only.
The first excited state is the spin-singlet formed in the dots B and C, plus one electron located in the dot A
\begin{equation}
\Psi_2 (\lambda=0) = \frac{ a^+_{C2} a^+_{B1} - a^+_{C1}a^+_{B2} }{\sqrt{2}} \times a^+_{A2} |0 \rangle  . \label{Psi2}
\end{equation}

In the first stage, it is necessary to form an initial state, the spin singlet, describing a chemical bond between dots A and B
(see Fig.1). We can do it starting at $\lambda = 0$ from the ground state (\ref{Psi1}) and proceeding to the excited state $\Psi_7
(\lambda=0.5)$.  We start with examining the response of the system to the sweep $V=E_P(N_C-N_B)$ where $\lambda$ is changed
linearly from 0 to 0.5 with various speed. The variations of the state energies with $\lambda$ are shown in Fig.~2. Several
avoided-level crossings can be clearly seen in this figure, and in the inset magnifying the region with many possible Landau-Zener
transitions. The reaction products are shown in Fig.~3 for various speeds of change. In the main panel, the initial state is the
ground one; while in the inset we start from the first excited state. At low speeds only the states near the ground state can be
occupied at the end of the evolution (adiabatic sweep). When $\dot{\lambda}$ increases, certain states can be populated while most
other states are almost empty. This subset of occupied states is controlled by the sequence of the avoided crossings and is unique
for a chosen parametric evolution (i.e., a chosen $H_0$ and $V$). Thus, one can simulate a desirable chemical/nuclear reaction by
changing $H_0$ and/or $V$, i.e., by changing the device architecture and gate structure, as well as by occupying certain initial
states. However, for chemical reactions with complicated potential energy surfaces the possible outcomes are numerous, and we can
adjust the function $\lambda(t)$ to follow a specific reaction pathway. Even for a fixed pathway, a variation of the speed
$\dot{\lambda}$ allows us to switch with a definite probability between the output states (i.e., output products). In particular,
it is evident from Fig.~3 that at the relatively fast sweep with almost 0.9 probability the final state of our reaction becomes
\begin{equation}
\Psi_7 (\lambda=0.5) = \frac{ a^+_{A2} a^+_{B1} - a^+_{A1}a^+_{B2} }{\sqrt{2}} \times a^+_{C2} |0 \rangle , \label{Psi7AB}
\end{equation}
which is the spin-singlet in the dots A and B, plus one electron located in dot C. It should be noted that this is essentially the
same state as $\Psi_2(\lambda=0)$, besides a change of the dot numeration.

In the second stage, we apply a very fast pulse, $ V_{\rm restore} = 0.5 E_P (N_A + N_B) - E_P N_C$, to restore the initial gate
potentials corresponding to the value $\lambda=0.$  After this pulse, the state of the system remains unchanged and the initial
state of the reaction $H + H_2 \rightarrow H_2 + H$ is formed, as shown in the left panel of Fig.~1.

In the third stage, we apply a pulse $V'=E_P(N_B-N_A)$, changing $\lambda$ from 0 to 0.5, to achieve the configuration shown in
the
 right panel of Fig.~1, However, this state,
\begin{equation}
\Psi_7' (\lambda =0.5) = \frac{ a^+_{C2} a^+_{A1} - a^+_{A2}a^+_{C1} }{\sqrt{2}} \times a^+_{B2} |0 \rangle  , \label{Psi7BC}
\end{equation}
can only be achieved with small probability $P=0.1$, even after a fast enough sweep (see inset of Fig.~3). To overcome this, we
apply the following {\it selective} sweep, using a step-like signal, $\lambda(t)=\dot{\lambda}_1 t$ for
$0<\lambda_1<\lambda<\lambda_2$ and $\lambda(t)=\dot{\lambda}_2 t$ overwise, with $\dot{\lambda}_1\gg \dot{\lambda}_2$. In this
case, we can activate Landau-Zener transitions at avoided crossings within a desirable interval $\lambda_1<\lambda<\lambda_2$.
Based on this technique we can mainly achieve a unique output quantum state (or product) \ref{Psi7BC} as shown in Fig.~4 for
$\lambda_1=0.3$ and $\lambda_2=0.325$ (blue circles). It is evident that the desired final state has a sufficiently large
probability $P=0.6$. However, this large probability can not be reached without carefully choosing $\lambda_1$ and $\lambda_2$.
Indeed, the red crosses in Fig.~4 show the standard final occupation probability.

In summary, we propose to model chemical reactions via electron redistributions between coupled semiconductor quantum dots. As an
example, the simplest chemical reaction, $H + H_2 \rightarrow H_2 + H$ is examined here with the three nuclei and three electrons
being simulated by the three-electron states in the triple-dot structure. We achieve the following bond redistribution between the
three dots after the following procedure: (i) starting from the ground state of the three-dot system which contains (with $P=1/3$)
a spin-triplet of the electron pair B==C, a fast adiabatic sweep is used to obtain the spin-singlet state for electrons between
dots A and B (A==B covalent bonding) with $P=0.9$; (ii) applying a sharp restoring pulse we return to the initial Hamiltonian,
without changing the state of the system; (iii) we apply a selective adiabatic sweep to the dots B and A and transfer the bond to
the covalent coupling between dots C and A, C==A (with $P=0.1$ with fast, but not selective sweep, and with $P=0.6$ with a slow
but selective sweep), which is described by the spin-singlet state of the electron pair which is shared by the dots A and C. In
stages (i) and (ii), the initial state (left panel of Fig.~1) is formed and in the stage (iii) it is transferred to the final
state (right panel of Fig.~1). It should be emphasized that the final state is obtained with non-unit probability, indicating
possible different reaction outcomes, as it is the case with real experiments \cite{HH2}. Complex chemical reactions are usually
characterized by a sophisticated potential energy surface. A fraction of this surface, containing a targeted simulation pathway,
 can be mapped (after rescaling) into the energy landscape of a quantum dot structure. Then, a fine tuning of the function
 $\lambda(t)$ can direct the artificial reaction along a certain chosen
pathway on the potential energy surface, with a distribution of the possible outcomes which reflects the results of real chemical
transformations. We believe that the direct emulation of chemical reactions with quantum electronic devices provides a new way in
analogous quantum computing which complements standard schemes of quantum state control and manipulation.

This work was supported in part by the National Security Agency, Laboratory of Physical Sciences, Army Research Office, National
Science Foundation grant No. EIA-0130383, and JSPS CTC Program. S.S. acknowledges support from the EPSRC ARF No. EP/D072581/1 and
AQDJJ network-programme. L.M. is partially supported by the NSF NIRT, grant ECS-0609146. A.S. is thankful to Sergei Studenikin,
Andy Sachrajda  and Louis Gaudreau for valuable discussions.

\newpage

\begin{figure}[ht]
\includegraphics[width=18.0cm]{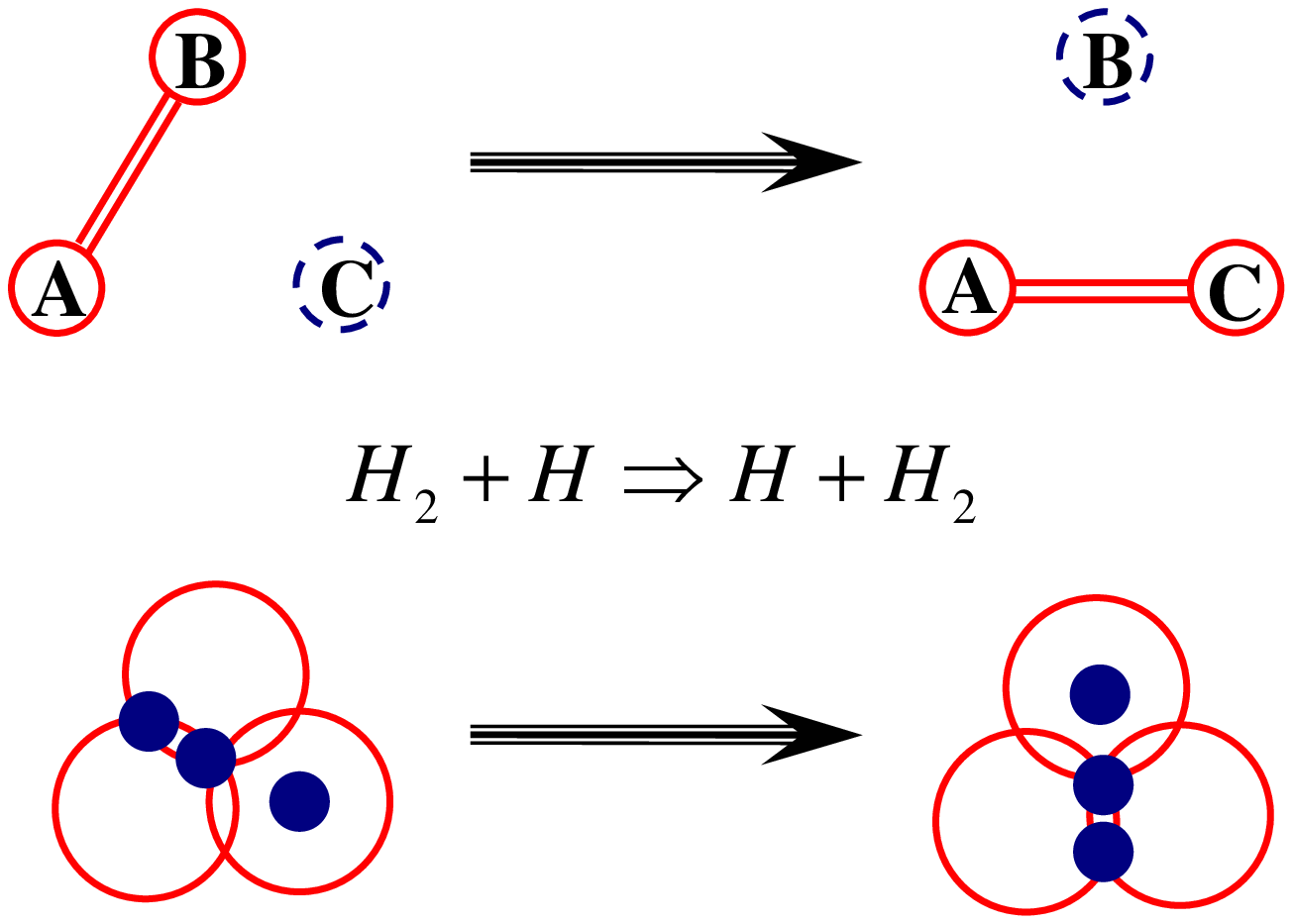}
\vspace*{12cm} \caption{\label{fig1} (Color online) Schematic diagram of the $H_2 + H \rightarrow H + H_2$ reaction showing the
bonds (upper panel) and the electron redistribution in the coupled quantum dot system (lower panel).}
\end{figure}

\begin{figure}
\includegraphics[width=18.0cm]{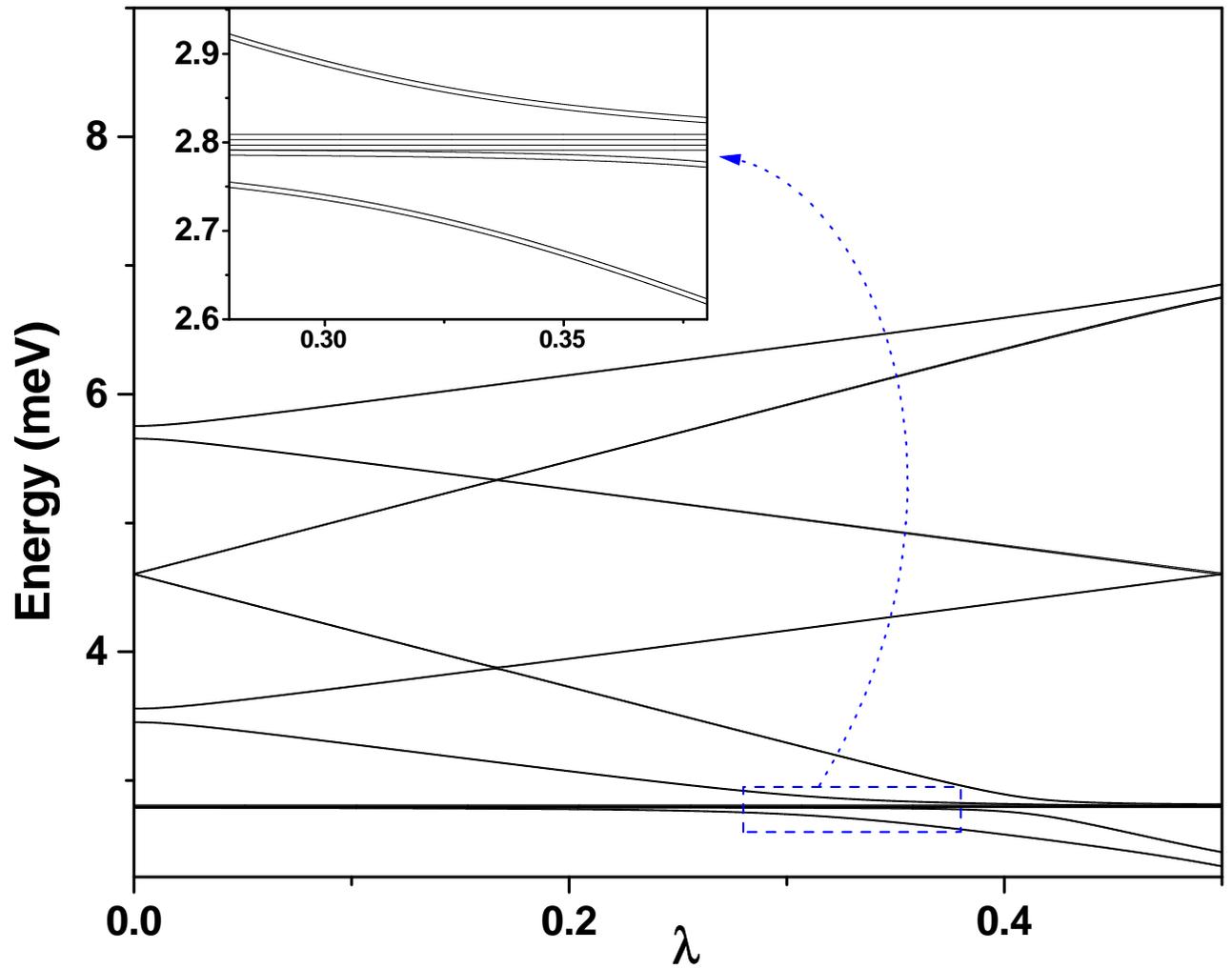}
\vspace*{14cm} \caption{Energies of the three-electron states in the triple-dot structure versus the parameter $\lambda$
representing different gate voltages. Inset: Magnified region with several avoided level crossings.}
\end{figure}

\begin{figure}
\includegraphics[width=18.0cm]{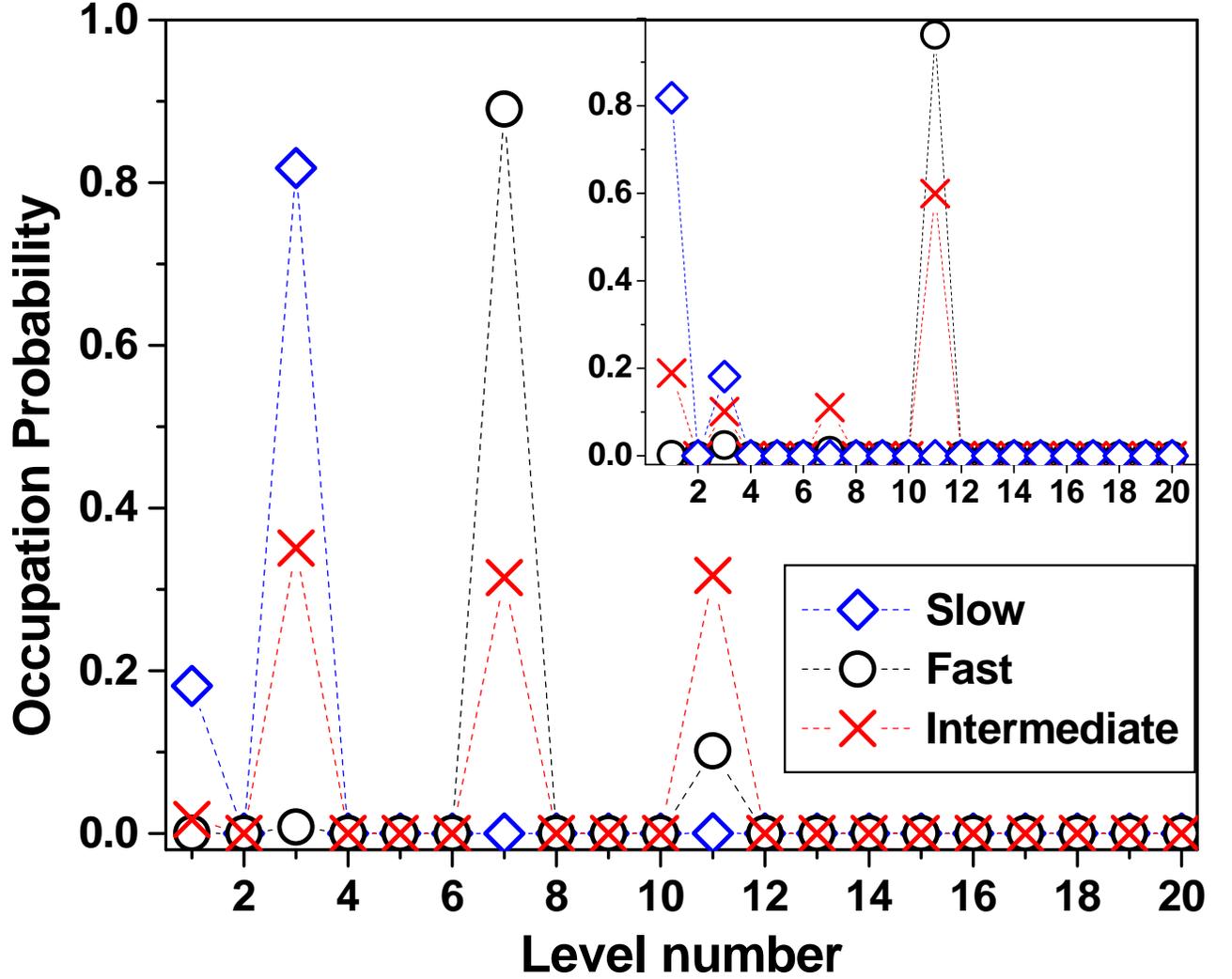}
\vspace*{14cm} \caption{(Color online) Level populations at the end of the parametric evolution of the Hamiltonian (\ref{eq_H0})
when starting from the ground (main panel) and the first excited (inset) states for different reaction speeds (in units of
meV/$\hbar$): $\dot{\lambda}=10^{-3}$ (blue diamonds), $10^{-2}$ (red crosses in the main panel), $7.5\cdot 10^{-2}$ (red crosses
in the inset), $10^{-1}$  (black circles in the main panel), 1 (black circles in the inset). Dashed lines connecting the symbols
are added as a guide to the eye.}
\end{figure}

\begin{figure}
\includegraphics[width=18.0cm]{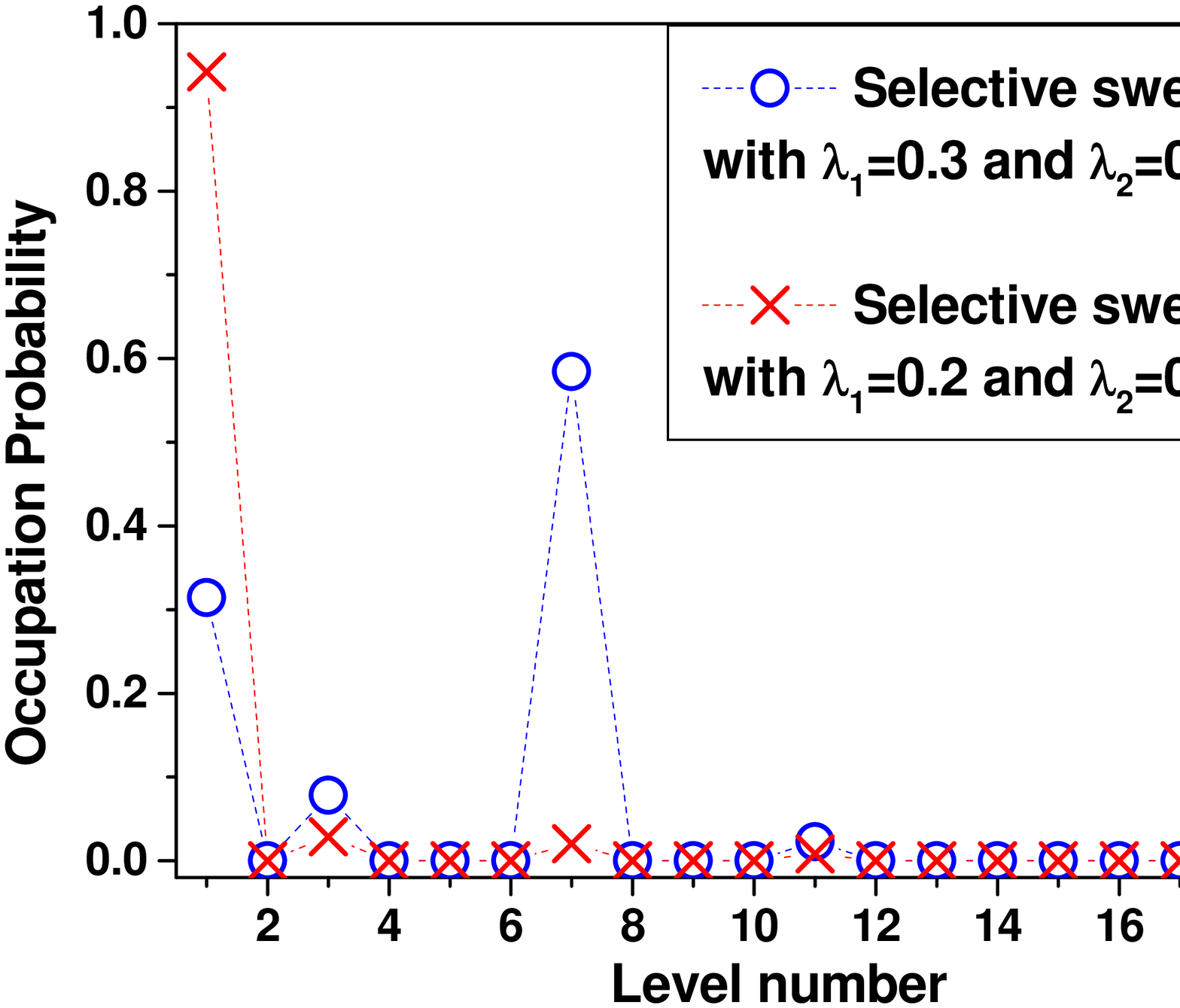}
\vspace*{14cm} \caption{(Color online) Occupation of states at the end of the parametric evolution when the step-like signal
$\lambda(t)$, described in the text, is applied with $\lambda_1=0.2$, $\lambda_2=0.225$ (red crosses) and $\lambda_1=0.3$,
$\lambda_2=0.325$ (blue circles). Note that output products corresponding to red crosses were found for almost all values of
$\lambda_1$ and $\lambda_2$, while output products shown by blue circles are unique and can be achieved only for a specific values
of $\lambda_1$ and $\lambda_2$.}
\end{figure}


\begin{references}

\bibitem{Lloyd1}
R.P.~Feynman, Int. J. Theor. Phys. {\bf 21}, 467 (1982); S.~Lloyd, Science {\bf 273}, 1073 (1996).

\bibitem{Manousakis1}
E.~Manousakis, J. Low Temp. Phys. {\bf 126}, 1501 (2002).

\bibitem{AspuruGuzik1}
A.~Aspuru-Guzik, A.D.~Dutoi, P.J.~Love, and M.~Head-Gordon, Science {\bf 309}, 1704 (2005).

\bibitem{1DExp}
M.~Kastner, Physics Today {\bf 46}, 24 (1993); S.~Tarucha, et al., Phys. Rev. Lett. {\bf 77}, 3613 (1996); M.~Stopa, Phys. Rev. B
{\bf 54}, 13767 (1996); L.P.~Kouwenhoven, et al., Science {\bf 278}, 1788 (1997).

\bibitem{Kouwenhoven2}
W.G.~van der Wiel, et al., Rev. Mod. Phys. {\bf 75}, 1 (2003).

\bibitem{DDExp}
See, e.g., A.W.~Holleitner, et al., Science {\bf 297}, 70 (2002); M.~Pioro-Ladriere, et al., Phys. Rev. Lett. {\bf 91}, 026803
(2003); M.~Rontani, et al., Phys. Rev. B {\bf 67}, 085327 (2004).

\bibitem{Barenco1}
A.~Barenco, et al., Phys. Rev. Lett. {\bf 74}, 4083 (1995).

\bibitem{Fujisawa1}
T.~Hayashi,et al., Phys. Rev. Lett. {\bf 91}, 226804 (2003); T.~Fujisawa, et al., Physica E {\bf 21}, 1046 (2004).

\bibitem{SEExp}
See, e.g., M.~Ciorga, et al., Phys. Rev. B {\bf 61}, R16315 (2000); A.C.~Johnson, et al., Nature {\bf 435}, 925 (2005);
A.K.~H\"uttel, et al., Phys. Rev. B {\bf 72}, 081310(R) (2005).

\bibitem{Field1}
M.~Field, et al., Phys. Rev. Lett. {\bf 70}, 1311 (1993).

\bibitem{TDExp}
A.~Vidan, et al., Appl. Phys. Lett. 85, 3602 (2004).

\bibitem{Louis06}
L.~Gaudreau, et al., Phys. Rev. Lett. {\bf 97}, 036807 (2006).

\bibitem{aqc} E.~Farhi et al., Science {\bf 292}, 472 (2001).

\bibitem{Zagoskin1}
 A.M. Zagoskin {et al.,} Phys. Rev. Lett. {\bf 98}, 120503
(2007).

\bibitem{Pechukas1983} P. Pechukas, Phys. Rev. Lett. {\bf 51}, 943 (1983).

\bibitem{HH2}
S.C.~Althorpe, et al., Nature {\bf 416}, 67 (2002); D.E.~Manolopoulos, Nature {\bf 419}, 266 (2002); S.A.~Harich, et al., Nature
{\bf 419}, 281 (2002).



\end{references}
\end{document}